\documentclass[aps,prc,twocolumn,floatfix,nofootinbib]{revtex4}
\usepackage{float,ulem}
\usepackage{graphicx,epsfig,epstopdf,amsmath,amssymb}
\usepackage{xcolor}

\newcommand{\be}{\begin{equation}}
\newcommand{\ee}{\end{equation}}
\newcommand{\ba}{\begin{eqnarray}}
\newcommand{\ea}{\end{eqnarray}}
\newcommand{\nn}{\nonumber\\}

\usepackage{scalerel}
\newcommand{\paral}{\stretchrel*{\parallel}{\perp}}

\usepackage{soul}

\begin{document}

\title{Causality and stability analysis of first-order field redefinition in relativistic hydrodynamics from kinetic theory}
\author{Sukanya Mitra}
\email{sukanya.mitra10@gmail.com}
\affiliation{Department of Nuclear and Atomic Physics, Tata Institute of Fundamental Research, Homi Bhabha Road, Mumbai 400005, India}

\begin{abstract}
In this work, the causality and stability of a first-order relativistic dissipative hydrodynamic theory, that redefines the hydrodynamic fields from a first principle microscopic estimation, 
have been analyzed. A generic approach of gradient expansion for solving the relativistic transport equation has been adopted using the Chapman-Enskog iterative method. Next, the momentum 
dependent relaxation time approximation (MDRTA) has been employed to quantify the collision term for analytical estimation of the field correction coefficients from kinetic theory. At linear 
regime, in local rest frame the dispersion relations are observed to produce a causal propagating mode. However, the acausality and instability reappear when a boosted background is considered 
for linear analysis. These facts point out relevant aspects regarding the methodology of extracting the causal and stable first order hydrodynamics from kinetic theory and indicate the appropriate 
approach to construct a valid first order theory with proper justification.
\end{abstract}
\maketitle

\section{Introduction}
The journey of relativistic dissipative hydrodynamic theory can be traced back from the relativistic extension of the Navier-Stokes~(NS) formalism introduced by Landau-Lifshitz~(LL)~\cite{LL} 
and Eckart~\cite{Eckart}. These theories are known as the first-order theories because of the presence of first order gradient corrections in the out of equilibrium deviations of the thermodynamic 
quantities such as entropy current. The problem occurs with these theories when they exhibit superluminal speed of signal propagation causing severe causality violation problem~\cite{Hiscock:1983zz}. 
This undesirable feature further associates instabilities within the system such that small departures of these fluids from equilibrium lead to rapid evolution away from equilibrium~\cite{Hiscock:1985zz}. 
These features pose major concern to the practical applicability of these theories  and make them unacceptable as a reasonable relativistic theory for fluids. 

To rescue the situation, second-order theories are introduced where the dissipative fluxes are promoted as the fundamental dynamical variables and give rise to relaxation type evolution equations. 
The second-order theory introduced by Israel and Stewart \cite{IS} known as Israel-Stewart (IS) theory, accepted as the standard theory of relativistic dissipative hydrodynamics, which has been shown 
to be both stable and causal in \cite{Hiscock:1983zz,Hiscock:1985zz,Hiscock:1987zz}. In \cite{Olson:1990rzl} the hyperbolicity of IS theory along with subluminal signal propagation has been demonstrated 
for linear perturbations around equilibrium. Since then, a range of second order theories like IS \cite{Muronga:2001zk,Muronga:2003ta} to recently developed DNMR \cite{Denicol:2012cn, Denicol:2012EPJA} 
and resummed BRSSS theory \cite{Baier:2007ix} have been introduced in the literature. In works like \cite{Denicol:2008ha,Pu:2009fj} the stability and causality have been analyzed for IS theory and in 
\cite{Brito:2020nou} the same  has been studied for DNMR theory. These studies give conditions involving the equation of state and the transport coefficients that ensure 
that these theories are indeed causal and stable. Proven to be free from causality and stability related issues at least for linear perturbations around equilibrium, they have been used for a wide range 
of hydrodynamic numerical simulations. Constraints to ensure causality for IS like theories in nonlinear, far-from-equilibrium regime have been recently explored in \cite{Bemfica:2019cop,Bemfica:2020xym}.

Recently, a new comprehensive formalism has been proposed by Bemfica, Disconzi, Noronha and Kovtun~(BDNK) to establish a causal and stable hydrodynamic theory~\cite{Bemfica:2017wps,Bemfica:2019knx,Kovtun:2019hdm} without incorporating extra dynamical degrees of freedom other than the fundamental ones such as temperature, hydrodynamic velocity and charge chemical potential. In other words, 
they derive first-order theories in the most general way possible that prohibit superluminal signal propagation as well as retain stability criteria besides other requirements 
like non-negative entropy production, that are essential for an acceptable hydrodynamic theory. The basic idea is to define the out of equilibrium thermodynamic variables in a 
general frame other than specified either by Landau-Lifshitz or Eckart, through their postulated constitutive relations.  This is the so called BDNK formalism, which proposed a class 
of stable and causal frames for the first order relativistic hydrodynamic theory.  In Refs.~\cite{Bemfica:2017wps,Hoult:2021gnb,Biswas:2022cla} the derivation of such a causal-stable first order theory
from relativistic Boltzmann equation has been studied in order to establish its kinetic theory origin.

Motivated from these studies, in this work, a first order theory has been derived, where the out of equilibrium thermodynamic fields are not uniquely defined and are subjected to 
include dissipative effects from the medium. Here, a first order relativistic dissipative hydrodynamic theory that includes the out of equilibrium contributions in thermodynamic 
fields purely from system interactions and recently been derived from relativistic transport equation by gradient expansion technique \cite{Mitra:2021owk}, has been employed to derive the 
dispersion equations and the associated modes. In order to linearize the non-trivial collision integral, momentum  dependent relaxation time approximation (MDRTA) has been adopted
for solving the relativistic transport equation as a model study \cite{Dusling:2009df,Teaney:2013gca,Kurkela:2017xis,Mitra:2020gdk,Rocha:2021lze,Rocha:2021zcw}. The obtained results
reveal interesting facts regarding the microscopic extraction of hydrodynamic field redefinition and its consequent effects on stability and causality of the theory. 

In the usual NS theory, the macroscopic thermodynamic quantities such as energy density and particle number density in the conservation equations are usually set to their equilibrium 
values even in the dissipative medium by imposing certain matching or fitting conditions. The resulting dispersion relation, at the limit of large wave number ($k$) gives rise to non 
propagating modes with $k^2$ dependence \cite{Romatschke:2009im,Denicol:2008ha}, identical to that of the diffusion process which is acausal with an infinite propagation speed. This behaviour 
($\omega (k)$ is growing faster than $k$) is a consequence of the acausal nature of the equations, which when linearized around equilibrium, the resulting modes become superluminal. 
The intention of the current analysis is to observe that if there are any changes in the linear modes after the first order field redefinition is introduced.  Here, first the equations 
of motion have been linearized around a hydrostatic equilibrium at local rest frame (LRF). Next, the equations have been tested with a more general state of equilibrium where the background is
Lorentz boosted with an arbitrary velocity. This generalization from zero to arbitrary background velocity is necessary, since this could result in a whole new set of modes. In this context,
the precedence of LL theory can be remembered. At zero chemical potential the LL theory gives two modes with zero background velocity, keeping the stability of the theory intact. It is the new 
mode that appears with a boosted background, drives the instability in the theory \cite{Denicol:2008ha}. A well defined relativistic theory cannot depend on whether one sets the background 
velocity to zero or not, and hence a consistency check between the boosted and non-boosted results is essential. 

The manuscript is organized as follows: In section~\ref{secII} the first-order relativistic hydrodynamics with thermodynamic field redefinition has been derived from relativistic transport 
equation of kinetic theory. In section~\ref{secIII} the dispersion relations and the modes are analyzed with a hydrostatic background at local rest frame giving asymptotic causality condition. 
Section~\ref{secIV} studies the same dispersion relations but with a Lorentz boosted background and demonstrates the additional acausal modes. Finally in section~\ref{secV} the work has been summarized 
with prior conclusions and useful remarks regarding the choice of hydrodynamic field redefinition and its consequence on the causality and stability of a first order theory. 

Throughout the manuscript I use natural unit~($\hbar = c = k_{B} = 1 $) and consider flat space-time with mostly negative metric signature $g^{\mu\nu} = \text{diag}\left(1,-1,-1,-1\right)$. The time-like 
fluid four velocity, $u^{\mu}$ satisfies the normalization condition $u^{\mu}u_{\mu}=1$. The projection operator orthogonal to $u^{\mu}$ is defined as $\Delta^{\mu\nu}= g^{\mu\nu}-u^{\mu}u^{\nu}$. The 
space-time partial derivative can be decomposed as $\partial_{\mu} = u_{\mu}D+\nabla_{\mu}$, with a temporal part $D=u^{\mu}\partial_{\mu}$ and a spatial part $\nabla_{\mu}=\Delta_{\mu\nu}\partial^{\nu}$. 
The traceless irreducible tensors of rank-1 and rank-2 are defined as $A_{\langle\mu\rangle}=\Delta_{\mu\nu}A^{\nu}$ and $ A_{\langle\mu} B_{\nu\rangle} =\Delta_{\mu\nu}^{\alpha\beta}A_{\alpha}B_{\beta}$, 
respectively with $\Delta_{\mu\nu}^{\alpha\beta}=\frac{1}{2}\left(\Delta_{\mu}^{\alpha}\Delta_{\nu}^{\beta}+\Delta_{\mu}^{\beta}\Delta_{\nu}^{\alpha}\right)-\frac{1}{3}\Delta_{\mu\nu}\Delta^{\alpha\beta}$. 

\section{Field redefinition in relativistic hydrodynamics}
\label{secII}
The basic problem is to estimate the first order out of equilibrium correction of the thermodynamic fields needed to define the particle four-flow $N^{\mu}$ and energy-momentum tensor 
$T^{\mu\nu}$. Here, relativistic transport equation serves the purpose by providing the first order correction in the single particle distribution function via gradient expansion technique. 
The first order Chapman-Enskog~(CE) gives the following integro-differential equation
over the the single particle distribution function~\cite{Degroot}, 
\be
p^{\mu}\partial_{\mu}f^{(0)}(x,p)=-\cal{L}[\phi]~.
\label{RTE}
\ee
Here, the first order particle distribution function is decomposed as $f=f^{(0)}+f^{(0)}(1\pm f^{(0)})\phi$ with $f^{(0)}=[\exp(\frac{p\cdot u}{T}-\frac{\mu}{T})\mp 1]^{-1}$ as the equilibrium
distribution for Bosons and Fermions respectively and $\phi$ denoting the distribution deviation from equilibrium. The linearized collision term~($\cal{L}[\phi]$) over the deviation of first 
order distribution function is given by,
\begin{align}
{\cal{L}}[\phi]=&\int d\Gamma_{p_1}d\Gamma_{p'}d\Gamma_{p'_1}f^{(0)}f_1^{(0)}\left(1\pm f'^{(0)}\right)\left(1\pm f_{1}'^{(0)}\right)\nn
&\left(\phi+\phi_1-\phi'-\phi'_1\right)W\left(p'p'_1|pp_1\right)~,
\label{coll}
\end{align}
with $d\Gamma_p=\frac{d^3p}{(2\pi)^3p^0}$ as the phase space factor and $W$ as the microscopic interaction rate. The equilibrium temperature, chemical potential and hydrodynamic four velocity 
of the system are denoted by $T, \mu$ and $u^{\mu}$, respectively.

In order to solve Eq.~\eqref{RTE}, I am adopting here one of the most conventional techniques. In transport equation~\eqref{RTE}, the time derivatives on the left hand side are eliminated by 
the spatial gradients using the first order thermodynamic identities such as, $\frac{DT}{T}=-\left(\frac{\partial P_0}{\partial\epsilon_0}\right)_{\rho_0}\left(\partial\cdot u\right)$, 
$D\tilde{\mu}=-\frac{1}{T}\left(\frac{\partial P_0}{\partial \rho_0}\right)_{\epsilon_0}\left(\partial\cdot u\right)$ and $\left(\epsilon_0+P_0\right)Du^{\mu}=\nabla^{\mu}P_0$. Here, 
$\rho_0,~\epsilon_0$ and $P_0$ are the equilibrium values of particle number density, energy density and hydrodynamic pressure of the system, respectively. It is to be noted, that the spatial 
gradients over field variables contribute to the thermodynamic forces and that is why in conventional methods to extract the single particle distribution function from transport equation, 
the time derivatives are eliminated by spatial gradients using first order thermodynamic identities.

Following this prescription, the left hand side of Eq.~\eqref{RTE} turns out to be a linear combination of thermodynamic forces as the following~\cite{Mitra:2015yaa},
\begin{align}
&f^{(0)}\left(1\pm f^{(0)}\right)
\bigg[\hat{Q}\partial\cdot u+\bigg(\frac{\tau_p}{\hat{h}}-1\bigg)\tilde{p}^{\mu}\nabla_{\mu}\tilde{\mu}+\tilde{p}^{\mu}\tilde{p}^{\nu}\sigma_{\mu\nu}\bigg]\nn
&=\frac{1}{T}\cal{L}[\phi]~,
\label{RTE1}
\end{align}
with,
$\hat{Q}=\frac{z^2}{3}+\tau_p^2((\frac{\partial P_0}{\partial\epsilon_0})_{{\rho}_0}-\frac{1}{3})+\tau_p\frac{1}{T}(\frac{\partial P_0}{\partial {\rho}_0})_{\epsilon_0}$.
$\sigma_{\mu\nu}=\nabla_{\langle{\mu}}u_{\nu\rangle}$ is the traceless, symmetric velocity gradient and $\hat{h}=(\epsilon_0+P_0)/{\rho}_0T$ is the scaled enthalpy per particle at equilibrium. 
The other used notations denote, $\tilde{p}^{\mu}={p^\mu}/{T}$ as the scaled particle 4-momenta, $\tau_p={(p\cdot u)}/{T}$ as the scaled particle energy at local rest frame,
$z=m/T$ as the scaled particle mass and $\tilde{\mu}={\mu}/T$.

Since the thermodynamic forces are independent, in order to be a solution of Eq.~\eqref{RTE1}, $\phi$ must be a linear combination of the thermodynamic forces as,
\be
\phi=A\left(\partial\cdot u\right)+B^{\nu}\nabla_{\nu}\tilde{\mu}+C^{\mu\nu}\sigma_{\mu\nu}~,
\label{phi}
\ee
with $B^{\mu}=B\tilde{p}^{\langle\mu\rangle}$ and $C^{\mu\nu}=C\tilde{p}^{\langle\mu}\tilde{p}^{\nu\rangle}$. It is customary to expand the unknown coefficients in the particle momentum basis such as, 
$A=\sum_{s=0}^p A^s(z,x) \tau_p^s,~ B=\sum_{s=0}^p B^s(z,x) \tau_p^s,~ C=\sum_{s=0}^p C^s(z,x) \tau_p^s$, with the series expanded up to any desired degree of accuracy.

The next job is to estimate the out of equilibrium dissipative correction in the thermodynamic fields. For this purpose, two most general field variables, namely the particle 4-flow ($N^{\mu}$) and 
the energy-momentum tensor ($T^{\mu\nu}$) is given respectively in their integral forms as the following,
\begin{align}
N^{\mu}=\int d\Gamma_p p^{\mu} f ~,~~~~ T^{\mu\nu}=\int d\Gamma_p p^{\mu}p^{\nu}f ~.
\label{field}
\end{align}
The out of equilibrium part of the distribution function $f$ in Eq.~\eqref{field} gives the necessary field corrections. The correction in particle number density ($\delta\rho$), energy density 
($\delta\epsilon$), pressure ($\delta P$), energy flow or momentum density ($W^{\alpha}$) and particle flux ($V^{\alpha}$) are given by,
\begin{align}
\delta \rho = &u_{\mu}\delta N^{\mu}=\int d\Gamma_pf^{(0)}\left(1\pm f^{(0)}\right)\left(p\cdot u\right)\phi~,
\label{rho}\\
\delta\epsilon=&u_\mu u_{\nu}\delta T^{\mu\nu}=\int d\Gamma_p f^{(0)}\left(1\pm f^{(0)}\right) \left(p\cdot u\right)^2\phi~,
\label{eps}\\
\delta P=&-\frac{1}{3}\Delta_{\mu\nu}\delta T^{\mu\nu}~,\nn
=&\frac{1}{3}\int d\Gamma_p f^{(0)}\left(1\pm f^{(0)}\right) \left[(p\cdot u)^2-m^2\right]\phi~,
\label{Pr}\\
W^{\mu}=&\Delta^{\alpha}_{\mu}u_{\nu}\delta T^{\mu\nu}~,\nn
=&\int d\Gamma_p f^{(0)}\left(1\pm f^{(0)}\right) p^{\langle\mu\rangle}\left(p\cdot u\right)\phi~,
\label{momflux}\\
V^{\mu}=&\Delta^{\alpha}_{\mu}N^{\mu}=\int d\Gamma_p f^{(0)}\left(1\pm f^{(0)}\right) p^{\langle\mu\rangle}\phi~.
\label{pnflux}
\end{align}
Here, $\delta N^{\mu}$ and $\delta T^{\mu\nu}$ are the first order dissipative corrections in particle four flow and energy momentum tensor, respectively.

Keeping upto first nonvanishing contribution from the collision operator, the respective corrections in the thermodynamic fields listed in Eqs.~\eqref{rho}-\eqref{pnflux}, are given by,
\begin{align}
&\delta\rho=c_{\Gamma}\left(\partial\cdot u\right),~~\delta\epsilon=c_{\Lambda}\left(\partial\cdot u\right),~~\delta P=c_{\Omega}\left(\partial\cdot u\right),\\
&W^{\alpha}=c_{\Sigma}\nabla^{\alpha}\tilde{\mu}~,~~V^{\alpha}=c_{\Xi}\nabla^{\alpha}\tilde{\mu}~,
\end{align}
with,
\begin{align}
c_{\Gamma}=&T(A^0a_1+A^1a_2+A^2a_3)~,
\label{field1}\\
c_{\Lambda}=&T^2(A^0a_2+A^1a_3+A^2a_4)~,
\label{field2}\\
c_{\Omega}=&\frac{T^2}{3}\big[A^0(a_2-z^2a_0)+A^1(a_3-z^2a_1)\nn
&+A^2(a_4-z^2a_2)\big]~,
\label{field3}\\
c_{\Sigma}=&T^2(B^0b_1+B^1b_2)~,
\label{field4}\\
c_{\Xi}=&T(B^0b_0+B^1b_1)~.
\label{field5}
\end{align}
The moment integrals are defined here as, 
$a_n=\int dF_p \tau_p^n,~\Delta^{\mu\nu}b_n=\int dF_p \tilde{p}^{\langle\mu\rangle}\tilde{p}^{\langle\nu\rangle}\tau_p^n,
~\Delta^{\alpha\beta\mu\nu}c_n=\int dF_p\tilde{p}^{\langle\mu}\tilde{p}^{\nu\rangle}\tilde{p}^{\langle\alpha}\tilde{p}^{\beta\rangle}\tau_p^n$, with $dF_p=d\Gamma_pf^{(0)}(1\pm f^{(0)})$.

It is to be noted here that, by the virtue of the collision integral properties ${\cal{L}}[p^{\mu}]=0$ and ${\cal{L}}[1]=0$ which follow from the energy-momentum and particle number conservation, 
the coefficients $A^0, A^1$ and $B^0$ can not be determined from the transport equation~\eqref{RTE1} and hence they are called the homogeneous solutions. Beyond that, $A^s, B^s$ and $C^s$ 
can be fully estimated from the transport equation and can be called the interaction solutions. In the present case they are estimated to be,
\begin{align}
&A^2=\frac{T}{[\tau_p^2,\tau_p^2]}\int dF_p\tau_p^2\hat{Q}~, \\
&B^1=\frac{T}{[\tau_p\tilde{p}^{\langle\mu\rangle},\tau_p\tilde{p}^{\langle\nu\rangle}]}\int dF_p\tilde{p}^{\langle\mu\rangle}\tilde{p}^{\langle\nu\rangle}\tau_p\bigg(\frac{\tau_p}{\hat{h}}-1\bigg)~,\\ 
&C^0=\frac{T}{[\tilde{p}^{\langle\alpha}p^{\beta\rangle},\tilde{p}^{\langle\mu}p^{\nu\rangle}]}\int dF_p\tilde{p}^{\langle\alpha}\tilde{p}^{\beta\rangle} \tilde{p}^{\langle\mu}\tilde{p}^{\nu\rangle}~.
\end{align}
The bracket quantities are defined as, $[\phi,\phi]=\int d\Gamma_p\phi {\cal{L}}[\phi]$ which are always non-negative. The homogeneous solutions are fully arbitrary and the field corrections in 
Eqs.~\eqref{field1}-\eqref{field5} due to them is attributed solely to the hydrodynamic frame choice. In certain situations the frame is so chosen that the homogeneous part exactly cancels the 
interaction part giving rise to field correction zero such that the field can be identified with its equilibrium value even in the dissipative medium. In current analysis, the non-equilibrium 
field corrections will be kept non-zero to generate the equations of motion that give rise to the dispersion relations and finally the frequency modes. 

However, these field corrections are not independent but constrained to give the dissipative flux of same tensorial rank. The coefficients are shown to follow,
\begin{align}
&c_{\Omega}-c_{\Lambda}\left(\frac{\partial P_0}{\partial\epsilon_0}\right)_{{\rho}_0}-c_{\Gamma}\left(\frac{\partial P_0}{\partial {\rho}_0}\right)_{\epsilon_0}=-\zeta~,
\label{coeffreln1}\\
&c_{\Sigma}-\hat{h}Tc_{\Xi}=-\frac{\lambda T}{\hat{h}}~,
\label{coeffreln2}
\end{align}
such that the field corrections add up to produce dissipative fluxes as,
\begin{align}
&\delta P-\left(\frac{\partial P_0}{\partial\epsilon_0}\right)_{{\rho}_0}\delta\epsilon-\left(\frac{\partial P_0}{\partial {\rho}_0}\right)_{\epsilon_0}\delta {\rho}=\Pi~,
\label{fieldreln1}\\
&W^{\alpha}-\hat{h}T V^{\mu}=q^{\alpha}~.
\label{fieldreln2}
\end{align}
Here, $\Pi=-T^2\int d\Gamma_p f^{(0)}(1\pm f^{(0)})\hat{Q}\phi=-\zeta(\partial\cdot u)$ and $q^{\alpha}=T^2\int d\Gamma_p f^{(0)}(1\pm f^{(0)}) \tilde{p}^{\langle\alpha\rangle}(\tau_p-\hat{h})\phi=-\frac{\lambda T}{\hat{h}}\nabla^{\alpha}\tilde{\mu}$
are respectively the first order bulk viscous and diffusion flow. The coefficient of bulk viscosity ($\zeta$) and thermal conductivity ($\lambda$) in this theory are respectively
given by,
\begin{align}
&\zeta=T^2\int d\Gamma_p f^{(0)}(1\pm f^{(0)})\hat{Q}A~,\\
&\lambda=-\frac{T}{3}\hat{h}\int d\Gamma_p f^{(0)}(1\pm f^{(0)})\tilde{p}^{\mu}\tilde{p}_{\mu}(\tau_p-\hat{h})B~.
\end{align}
Now since, $\hat{Q}f^{(0)}(1\pm f^{(0)})=\frac{1}{T}{\cal{L}}[A]$ and 
$\tilde{p}^{\langle\mu\rangle}\big(\frac{\tau_p}{\hat{h}}-1\big)f^{(0)}(1\pm f^{(0)})=\frac{1}{T}{\cal{L}}[B^{\mu}]$, then by virtue of self adjoint property of collision integral
$\int d\Gamma_p\psi{\cal{L}[\phi]}=\int d\Gamma_p\phi{\cal{L}[\psi]}$ with $\psi=\psi(x,p^{\mu})$, $\zeta$ and $\lambda$ do not include the homogeneous solutions and purely depends
upon interactions. Equations~\eqref{coeffreln1} and \eqref{coeffreln2} reveal that these combinations are frame invariant as suggested by Ref.~\cite{Kovtun:2019hdm} which retain only the 
interaction part of the field corrections through the physical transport coefficients associated with dissipative fluxes. Detailed discussion of this derivation for any order of gradient expansion will be available 
in Ref.~\cite{Mitra:2021owk}.
Including field corrections, the expressions for particle four-flow and energy-momentum tensor are respectively given as follows,
\ba
N^{\mu}=&&({\rho}_0+\delta {\rho})u^{\mu}+V^{\mu}~,
\label{numberflow}\\
T^{\mu\nu}=&&(\epsilon_0+\delta\epsilon)u^{\mu}u^{\nu}-(P_0+\delta P)\Delta^{\mu\nu}\nn
&&+(W^{\mu}u^{\nu}+W^{\nu}u^{\mu})+\pi^{\mu\nu}~.
\label{enmomflow}
\ea
Here, $\pi^{\mu\nu}=\Delta^{\mu\nu}_{\alpha\beta}\delta T^{\alpha\beta}=2\eta\sigma^{\mu\nu}$ is the first order shear stress tensor with $\eta$ as the shear viscous coefficient.

The next job is to implement a microscopic model that can explicitly determine the field correction coefficients from Eqs.~\eqref{field1}-\eqref{field5}. Since, the homogeneous  part in the 
field correction of Eqs.~\eqref{field1}-\eqref{field5} can be chosen arbitrarily, here I am considering only the interaction correction provided by the transport equation itself. For the same, 
I propose here solving the relativistic transport equation~\eqref{RTE} in momentum dependent relaxation time approximation (MDRTA). The idea is just to replace ${\cal{L}}[\phi]$ in Eq.~\eqref{RTE} 
with the help of relaxation time $\tau_R$ of single particle distribution function as follows, 
\be
\tilde{p}^{\mu}\partial_{\mu}f=-\frac{\tau_p}{\tau_R}f^{(0)}\left(1\pm f^{(0)}\right)\phi ~,~~~\tau_R(x,p)=\tau_R^0(x) \tau_p^n,
\label{Relax}
\ee
where the momentum dependence of $\tau_R$ is expressed as a power law of the scaled particle energy $\tau_p$ in comoving frame, with $\tau_R^0$ as the momentum independent part  
and $n$ as the exponent specifying the power of the scaled energy. 
In Refs.~\cite{Mitra:2021owk,Mitra:2020gdk} the interaction part of the out of equilibrium 
field corrections have been estimated using the MDRTA technique from the relativistic transport equation. Here, the first order field correction coefficients are listed below,
\begin{align}
&\frac{c_{\Lambda}}{\tau_R^0}=
T^2\bigg[\frac{z^2}{3}a_{n+1}+\left\{\left(\frac{\partial P_0}{\partial \epsilon_0}\right)_{{\rho}_0}-\frac{1}{3}\right\}a_{n+3}\nn
&~~~~~~~~~~+\frac{1}{T}\left(\frac{\partial P_0}{\partial {\rho}_0}\right)_{\epsilon_0} a_{n+2}\bigg],
\label{coeff11}\\
&\frac{c_{\Gamma}}{\tau_R^0}=T\bigg[
\frac{z^2}{3}a_{n}+\left\{\left(\frac{\partial P_0}{\partial \epsilon_0}\right)_{{\rho}_0}-\frac{1}{3}\right\}a_{n+2}\nn
&~~~~~~~~~~~~~~+\frac{1}{T}\big(\frac{\partial P_0}{\partial {\rho}_0}\big)_{\epsilon_0} a_{n+1}\bigg],
\label{coeff12}\\
&\frac{c_{\Omega}}{\tau_R^0}=T^2\bigg[\frac{z^2}{9}a_{n+1}+\frac{1}{3}\left\{\left(\frac{\partial P_0}{\partial \epsilon_0}\right)_{{\rho}_0}-\frac{1}{3}\right\}a_{n+3}\nn
&~~~~~~+\frac{1}{3T}\left(\frac{\partial P_0}{\partial {\rho}_0}\right)_{\epsilon_0} a_{n+2}-\frac{z^4}{9}a_{n-1}\nn
&~~~~~~-\frac{z^2}{3}\left\{\left(\frac{\partial P_0}{\partial \epsilon_0}\right)_{{\rho}_0}-\frac{1}{3}\right\}a_{n+1}-\frac{z^2}{3T}\left(\frac{\partial P_0}{\partial {\rho}_0}\right)_{\epsilon_0} a_n\bigg],
\label{coeff13}\\
&\frac{c_{\Sigma}}{\tau_R^0}=T^2\left[\frac{1}{\hat{h}}b_{n+1}-b_{n}\right]~,
\label{coeff14}\\ 
&\frac{c_{\Xi}}{\tau_R^0}=T\left[\frac{1}{\hat{h}}b_{n}-b_{n-1}\right]~.
\label{coeff15}
\end{align}
The corresponding first order transport coefficients bulk viscosity ($\zeta$), thermal conductivity ($\lambda$) and shear viscosity ($\eta$) in MDRTA
are given by,
\begin{align}
&\frac{\zeta}{T^2\tau_R^0}=\frac{z^4}{9}a_{n-1}+\bigg\{\big(\frac{\partial P_0}{\partial \epsilon_0}\big)_{{\rho}_0}-\frac{1}{3}\bigg\}^2a_{n+3}\nn
&+\frac{2z^2}{3T}\big(\frac{\partial P_0}{\partial {\rho}_0}\big)_{\epsilon_0} a_n+\frac{2}{T}\bigg\{\big(\frac{\partial P_0}{\partial \epsilon_0}\big)_{{\rho}_0}-
\frac{1}{3}\bigg\}\big(\frac{\partial P_0}{\partial {\rho}_0}\big)_{\epsilon_0} a_{n+2}\nn
&+\frac{1}{T^2}\big(\frac{\partial P_0}{\partial {\rho}_0}\big)^2_{\epsilon_0} a_{n+1}+\frac{2z^2}{3}\bigg\{\big(\frac{\partial P_0}{\partial \epsilon_0}\big)_{{\rho}_0}-\frac{1}{3}\bigg\}a_{n+1}~,
\label{zeta}\\
&\frac{\lambda T}{T^2\tau_R^0}=-\big\{b_{n+1}-2\hat{h}b_n+\hat{h}^2b_{n-1}\big\}~, 
\label{lambda}\\
&\frac{\eta}{T^2\tau_R^0}=\frac{1}{2}c_{n-1}~.
\label{eta}
\end{align}
The conservation of particle four-flow and energy-momentum tensor along with the non-negativity of entropy production rate have been confirmed within the present theory in Ref.~\cite{Mitra:2021owk}.

\section{Causality and stability analysis in Local Rest Frame}
\label{secIII}
To analyze the modes, first small perturbations of the hydrodynamic variables are considered around a hydrostatic equilibrium state of the fluid which is in local 
rest frame such as, 
\ba
&T=T_0+\delta T(t,x)~,~~~\tilde{\mu}=\tilde{\mu}_0+\delta\tilde{\mu}(t,x)~,\nn
&u^{\mu}=(1,\vec{0})+\delta u^{\mu}(t,x)~.
\ea
In linear approximation, the velocity perturbation has only spatial components $\delta u^{\mu}=(0,\delta u^x,\delta u^y,\delta u^z)$,
since one needs $u_0^{\mu}\delta u_{\mu}=0$ to retain the normalization condition.
It is convenient to express these fluctuations in their plane wave solutions via a Fourier transformation $\delta\psi(t,x)\rightarrow e^{i(\omega t-kx)} \delta\psi(\omega,k)$,
with wave 4-vector $k^{\mu}=(\omega,k,0,0)$. Following this prescription, the conservation equations $\partial_{\mu}N^{\mu}=0$ and $\partial_{\mu}T^{\mu\nu}=0$, over the Eqs.~\eqref{numberflow} and \eqref{enmomflow} give the dispersion relations. Following the convention of~\cite{Baier:2007ix}, retaining the component of $\delta u^{\mu}$ parallel to
$k^{\mu}$, the dispersion relation for the longitudinal or sound mode is obtained as the following,
\be
\omega^3(1+Ak^2)-iB\omega^2 k^2-\omega(Ck^2+Dk^4)+iEk^4=0~,
\label{dispersion}
\ee
with,
\begin{align}
A&=\hat{h}\tilde{c}_{\Sigma}(\tilde{c}_{\Lambda}-\tilde{c}_{\Gamma})~,
\label{coeffA}\\
B&=(4\eta/3+\zeta+\lambda T)/(\epsilon_0+P_0)~,
\label{coeffB}\\
C&=c_s^2~,\label{coeffC}\\
D&=(4\eta/3+\zeta)\lambda T/(\epsilon_0+P_0)^2\nn
&+\hat{h}(\tilde{c}_{\Lambda}-\tilde{c}_{\Gamma})
\bigg[\big(\frac{\partial P_0}{\partial\epsilon_0}\big)_{\rho_0}\tilde{c}_{\Sigma}+\frac{1}{\hat{h}}\frac{1}{T}\big(\frac{\partial P_0}{\partial\rho_0}\big)_{\epsilon_0}\tilde{c}_{\Xi}\bigg]~,
\label{coeffD}\\
E&=c_s^2\lambda T/(\epsilon_0+P_0)~.
\label{coeffE}
\end{align}
The used notations read, $\tilde{c}_{\Lambda}=c_{\Lambda}/(\epsilon_0+P_0),~\tilde{c}_{\Sigma}=c_{\Sigma}/(\epsilon_0+P_0),~\tilde{c}_{\Gamma}=c_{\Gamma}/\rho_0,~\tilde{c}_{\Xi}=c_{\Xi}/\rho_0$
and $c_s^2=\big(\frac{\partial P_0}{\partial\epsilon_0}\big)_{\rho_0}+\frac{1}{\hat{h}}\frac{1}{T}\big(\frac{\partial P_0}{\partial\rho_0}\big)_{\epsilon_0}$ is the velocity of sound squared 
\cite{Monnai:2012jc}. The coefficients $B, C, E$ and first part of $D$ being the function of physical transport coefficients associated with dissipative fluxes only (which are independent of hydrodynamic 
field corrections), they will be present in the usual NS theory as well, i.e. without field redefinition in out of equilibrium scenario. However, as mentioned earlier, in most of the studies $c_{\Lambda}$
and $c_{\Gamma}$ are set to zero employing certain frame choice in order to keep the energy density and particle number density at their equilibrium values even in dissipative medium. In such 
cases, $A$ and the second part of $D$ in Eq.~\eqref{dispersion} vanish. In such situations, propagating modes appears only at small $k$ values with a propagation speed of usual sound velocity $c_s$.
The problem occurs at large $k$ limit where the propagating modes are changed to non-propagating modes with $k^2$ dependence which indicate acausality~\cite{Denicol:2008ha}.
Here the sound channel dispersion relation Eq.~\eqref{dispersion} is analyzed in presence of all the field corrections. 

At small $k$ values the dispersion relation gives,
\begin{align}
&\omega^{\paral}_{1,2}=\frac{i}{2}\left[\frac{4\eta/3+\zeta}{(\epsilon_0+P_0)}\right]k^2\pm k c_s~,
\label{mode11}\\
&\omega^{\paral}_3=i \left[\frac{\lambda T}{(\epsilon_0+P_0)}\right] k^2~,
\label{mode12}
\end{align}
which is identical to the usual NS theory without field redefinition. $\omega^{\paral}_{1,2}$ is the conventional propagating sound mode with a propagation velocity of the speed of sound. 
$\omega^{\paral}_{3}$ is the purely non-propagating heat-diffusion mode. The imaginary part of all the modes being always positive by the virtue of positive physical transport coefficients,
the modes are always stable.

It is the large $k$ limit that differs from the conventional NS theory. At large $k$, the dispersion relation renders,
\begin{align}
&\omega^{\paral}_{1,2}=\frac{i}{2}\bigg\{\frac{B}{A}-\frac{E}{D}\bigg\}\pm k\sqrt{\frac{D}{A}}~,
\label{mode21}\\
&\omega^{\paral}_3=i\frac{E}{D}~,
\label{mode22}
\end{align}
where positive values of $D/A$ give two propagating modes via the real part of frequency $\omega^{\paral}_{1,2}$. 
Eq.~\eqref{coeffA} shows that in absence of field redefinition in energy density and particle number density, $A$ (as well as the second term of $D$) vanishes and consequently Eq.~\eqref{dispersion} produces 
only the non-propagating modes at large $k$ limit. Since propagation speed of the fluid is characterized by the group velocity of the propagating mode, in order to analyze the causality of the mode,
here the asymptotic value of group velocity $(v_g)$ has been defined as follows,
\be
v_g=\lim_{k\rightarrow\infty}\bigg\vert\frac{\partial \textrm{~Re}({\omega})}{\partial k}\bigg\vert=\sqrt\frac{D}{A}~.
\label{vg1}
\ee
In order to be subluminal, the theory must satisfy $D/A<1$ along with $D/A>0$. Eqs.~\eqref{coeffA} and~\eqref{coeffD} show that $A$ and $D$ explicitly depend upon the field correction coefficients. 
So it can be derived that in order to preserve causality of the propagating mode at local rest frame, the coefficients must satisfy the following relation,
\begin{align}
&\frac{(4\eta/3+\zeta)\lambda T/(\epsilon_0+P_0)^2}{\hat{h}\tilde{c}_{\Sigma}(\tilde{c}_{\Lambda}-\tilde{c}_{\Gamma})}<\nn
&\bigg[1-\bigg\{\big(\frac{\partial P_0}{\partial\epsilon_0}\big)_{\rho_0}+\frac{1}{\hat{h}}\frac{1}{T}\big(\frac{\partial P_0}{\partial\rho_0}\big)_{\epsilon_0}\frac{\tilde{c}_{\Xi}}{\tilde{c}_{\Sigma}}\bigg\}\bigg].
\label{vg2}
\end{align}
Equation~\eqref{vg2} is the asymptotic causality condition of the theory at local rest frame. Under the MDRTA formalism it can be shown that with small mass (typically $z$ values below $0.25$) and non-zero values of the 
exponent $n$ in Eq.~\eqref{Relax} (particularly on the negative $n$ side), the asymptotic causality condition $0<v_g<1$ is indeed satisfied with linearized perturbations around LRF equilibrium. 

The causal propagating mode of Eq.~\eqref{mode21} with thermodynamic field redefinition is however interesting, but certainly not conclusive for verifying the causality of the theory
as a whole. Since local rest frame is not the most general equilibrium state, it is crucial to check also the situation with a more general state of equilibrium, i.e., to consider 
linear disturbances around the background with hydrodynamic velocity $u^{\mu}_0=\gamma(1,\textbf{v})$, where the velocity $\textbf{v}$ is nonzero and $\gamma = 1/\sqrt{1-\textbf{v}^2}$.
Anyway, with field redefinition
and at LRF background, the shear channel does not improve and gives the same mode as in NS theory, $\omega^{\perp}=i[\eta/(\epsilon_0+P_0)]k^2$. Though this mode is stable, it is certainly
not causal. That is why, in order to have a more rigorous study, in the next section a more general equilibrium with a boosted background will be considered for linear stability and causality analysis.

\section{Causality and Stability  analysis in Lorentz-Boosted frame}
\label{secIV}

The background fluid is now considered to be boosted along x-axis with a constant velocity $\textbf{v}$, $u^{\mu}_0=\gamma(1,\textbf{v},0,0)$. The corresponding velocity fluctuation
is $\delta u^{\mu}=(\gamma \textbf{v} \delta u^x,\gamma\delta u^{x},\delta u^y,\delta u^z)$ which again gives $u^{\mu}_0\delta u_{\mu}=0$ to maintain velocity normalization.
The dispersion relations can be obtained in the boosted frame by giving the transformations, $\omega \rightarrow \gamma(\omega - k \textbf{v})$ and $k^2\rightarrow \gamma^2(\omega-kv)^2-\omega^2+k^2$ 
to the local rest frame~\cite{Bemfica:2019knx}.

The dispersion relation for the shear channel with boosted background turns out to be a quadratic equation of $\omega$. Here, I address the shear modes in two limiting cases.

At small $k$ limit, the shear modes are,
\begin{align}
&\omega^{\perp}_1=\textbf{v} k + {\cal{O}}(k^2)~,
\label{shear1}\\
&\omega^{\perp}_2=-\frac{i}{\gamma \Gamma  \textbf{v}^2}+\frac{(2-\textbf{v}^2)}{\textbf{v}}k+{\cal{O}}(k^2)~,
\label{shear2}
\end{align}
with $\Gamma=\eta/(\epsilon_0+P_0)$.
In small $k$ limit, it is clear that the mode $\omega^{\perp}_1$ is a propagating mode with just the background velocity itself. The imaginary part of the other shear mode $\omega^{\perp}_2$,  
is always negative since $\eta$ is a positive quantity, indicating the mode is unstable. For a background velocity $0<\textbf{v}<1$, the mode is acausal as well.

At large $k$ limit, the shear modes becomes,
\be
\omega^{\perp}_{1,2}=\frac{1}{\textbf{v}}k~,
\ee
which can be readily seen as acausal for any acceptable background velocity $\textbf{v}$. So, with a boosted background velocity, the causality and stability both are violated in the shear channel
of the fluid.

With boosted background, the dispersion relation for the sound channel becomes an extremely complicated fifth order polynomial which is not possible to solve analytically. For this difficulty, here 
I again address the numerical solution of the modes in two limiting cases.

In small $k$ limit, the sound modes becomes,
\begin{align}
&\omega^{\paral}_1=\textbf{v} k+{\cal{O}}(k^2)~,
\label{sound1}\\
&\omega^{\paral}_{2,3}=\frac{1}{2}\left[M\pm \sqrt{M^2-4N}\right]k+{\cal{O}}(k^2)~,
\label{sound2}\\
&\omega^{\paral}_{4,5}=\frac{i}{2}\left[Q\pm \sqrt{Q^2+4R}\right]+{\cal{O}}(k)~,
\label{sound3}
\end{align}
with,
\begin{align}
&M=\frac{2\textbf{v}(c_s^2-1)}{(c_s^2 \textbf{v}^2-1)}~,~~~~N=\frac{c_s^2-\textbf{v}^2}{(c_s^2 \textbf{v}^2-1)}~,\\
&Q=\frac{E\textbf{v}^2-B}{\gamma(D\textbf{v}^2-A)}~,~~~~R=\frac{c_s^2 \textbf{v}^2 -1}{\gamma^2 \textbf{v}^2(D\textbf{v}^2-A)}~.
\end{align}
Here, the coefficients $A,B,D,E$ are listed in Eqs.~\eqref{coeffA}-\eqref{coeffE}.
The ${\cal{O}}(k^2)$ terms of $\omega^{\paral}_1$ and $\omega^{\paral}_{2,3}$ are complicated function of field correction coefficients which at $\textbf{v}\rightarrow 0$ reduce to the same non-propagating
parts of the LRF modes of Eqs.~\eqref{mode12} and~\eqref{mode11}, respectively. So with that, at vanishing background velocity, the modes in Eqs.~\eqref{sound1} and~\eqref{sound2} boil down to LRF
sound modes with small wave number. It is the $\omega^{\paral}_{4,5}$ modes which were not present in the local rest frame. It has been tested that no combination of the field correction 
coefficients can produce a positive imaginary part for $\omega^{||}_{4,5}$ and the modes become unstable.

In the limit of large wave numbers, an expansion of the form $\omega^{\paral}=v_g^{\paral}k+\sum_{n=0}^{\infty}c_nk^{-n}$ can be used a solution~ \cite{Brito:2020nou}. The roots of $v_g^{\paral}$
are obtained as,
\begin{align}
&v_{g,1}^{\paral}=\textbf{v}~,\\
&v_{g,2,3}^{\paral}=\frac{\left[\textbf{v}(A-D)\pm \sqrt{AD-2AD\textbf{v}^2+AD\textbf{v}^4}\right]}{A-D\textbf{v}^2}~,\\
&v_{g,4,5}^{\paral}=\pm \frac{1}{\textbf{v}}~.
\end{align}
At $\textbf{v}\rightarrow 0$, $v_{g,1}^{\paral}$ vanishes and $v_{g,2,3}^{\paral}$ reduce to Eq.~\eqref{vg1} with asymptotic group velocity $v_g=\sqrt{D/A}$ of local rest frame.
With $0<\textbf{v}<1$, these three modes remain always subluminal as long as the causal parameters of the field correction coefficients are being used from condition Eq.~\eqref{vg2}. 
It is the two new roots $v_{g,4,5}^{\paral}$~, that are always acausal for $0<\textbf{v}<1$. So finally it can be concluded that, although at local rest frame the asymptotic causality
condition and stability criteria are maintained, the new modes of shear and sound channels due to the boosted background, are conclusively showing that the theory is acausal 
and unstable, irrespective of whatever values of the field correction coefficients are taken.

\section{Summary and conclusion}
\label{secV}
In this work, I analyze the causality and stability of a relativistic hydrodynamic theory, including the out of equilibrium field redefinition estimated from relativistic kinetic equation.
In local rest frame, when linearized around an equilibrium, the equations of motion give a propagating mode which was previously absent for the usual NS theory, along with the asymptotic 
causality condition~[Eq.~\eqref{vg2}] obeyed for certain constrained values of the field correction coefficients. However, when the background fluid is boosted with an arbitrary velocity 
$\textbf{v}$, it is observed
that new modes are appearing on the top of the LRF modes, which are both acausal and unstable. This observation reveals two important points here. The first one is quite straightforward.
In order to analyze the causality and stability of a theory, observing Fourier modes in the local rest frame is not only insufficient, but sometimes can be misleading (like 
the present case) as well. In order to have the correct conclusions, the most general equilibrium state is needed to be implemented. In fact causality is a general property of the equations of motion, 
and the ability of doing a Fourier analysis is not requisite. Considering the fact, in future a full non linear analysis with thermodynamic field redefinition is in order
to study the causality and stability of the theory in a more general way. 

The second point is somewhat more significant. The derivation of a first order theory, introducing non-equilibrium field corrections in fundamental macroscopic quantities
is a recent venture which is establishing itself as an authentic framework for the relativistic hydrodynamics. In their analysis, the contribution from the homogeneous part of the
non-equilibrium distribution function is attributed to generate new terms that give rise to causality and stability of the theory \cite{Bemfica:2017wps}. Motivated by these works, 
in the current analysis the effects of the purely interacting or inhomogeneous contribution from the distribution function have been tested in the hydrodynamic field corrections. In order to do that, 
the most general approach of gradient expansion, the Chapman-Enskog method has been used that expresses the out of equilibrium distribution function purely in terms of spatial gradients 
(eliminating the time derivatives imposing conservation equations). The resulting field corrections lack the time derivatives of the general constitutive relations of BDNK formalism and 
only include the spatial gradients resembling thermodynamic forces. The theory turns out to be causal and stable at local rest frame background, however it shows anomaly 
in a more general boosted background. This is the limitation of general Chapman-Enskog methodology which replaces the time derivatives with pure spatial gradients. 
The presence of terms including comoving derivatives is fundamental for causality and stability in any first ­order formulation. So, an alternate 
microscopic approach of solving the transport equation is required \cite{Biswas:2022cla} that retain the comoving derivatives, because this work singularly proves that they are the crucial 
counterparts in the hydrodynamic field corrections that definitively make the theory causal and stable.
\\
\\
\begin{acknowledgments}
Author duly thanks Rajesh Biswas for the numerical analysis, useful discussions and critical reading of the manuscript. Author acknowledges funding support from DNAP, TIFR, INDIA. Author 
also thanks the anonymous referee for valuable feedback that helped to improved the manuscript.
\end{acknowledgments}


\begin{thebibliography}{99}

\bibitem{LL}
L.D. Landau and E.M. Lifshitz, Fluid Mechanics (Elsevier, Amsterdam,1987).

\bibitem{Eckart}
C. Eckart, Phys. Rev. \textbf{58} (1940, )919.

\bibitem{Hiscock:1983zz}
W.~A.~Hiscock and L.~Lindblom,
Annals Phys. \textbf{151} (1983), 466-496.

\bibitem{Hiscock:1985zz}
W.~A.~Hiscock and L.~Lindblom,
Phys. Rev. D \textbf{31} (1985), 725-733.

\bibitem{Hiscock:1987zz}
W.~A.~Hiscock and L.~Lindblom,
Phys. Rev. D \textbf{35} (1987), 3723-3732.

\bibitem{IS}
Israel W 1976 Ann. Phys., NY 100 310, Israel W and Stewart J M 1979 Ann. Phys., NY 118 341.

\bibitem{Olson:1990rzl}
T.~S.~Olson,
Annals Phys. \textbf{199} (1990), 18.

\bibitem{Muronga:2001zk}
A.~Muronga,
Phys. Rev. Lett. \textbf{88} (2002), 062302.

\bibitem{Muronga:2003ta}
A.~Muronga,
Phys. Rev. C \textbf{69} (2004), 034903.

\bibitem{Denicol:2012cn}
G.~S.~Denicol, H.~Niemi, E.~Molnar and D.~H.~Rischke,
Phys. Rev. D \textbf{85} (2012), 114047.

\bibitem{Denicol:2012EPJA}
G.~S.~Denicol, E.~Moln\'ar, H.~Niemi and D.~H.~Rischke,
Eur. Phys. J. A \textbf{48} (2012), 170.

\bibitem{Baier:2007ix}
R.~Baier, P.~Romatschke, D.~T.~Son, A.~O.~Starinets and M.~A.~Stephanov,
JHEP \textbf{04} (2008), 100.

\bibitem{Denicol:2008ha}
G.~S.~Denicol, T.~Kodama, T.~Koide and P.~Mota,
J. Phys. G \textbf{35} (2008), 115102.

\bibitem{Pu:2009fj}
S.~Pu, T.~Koide and D.~H.~Rischke,
Phys. Rev. D \textbf{81} (2010), 114039.

\bibitem{Brito:2020nou}
C.~V.~Brito and G.~S.~Denicol,
Phys. Rev. D \textbf{102} (2020) no.11, 116009.

\bibitem{Bemfica:2019cop}
F.~S.~Bemfica, M.~M.~Disconzi and J.~Noronha,
Phys. Rev. Lett. \textbf{122} (2019) no.22, 221602.

\bibitem{Bemfica:2020xym}
F.~S.~Bemfica, M.~M.~Disconzi, V.~Hoang, J.~Noronha and M.~Radosz,
Phys. Rev. Lett. \textbf{126} (2021) no.22, 222301.

\bibitem{Bemfica:2017wps}
F.~S.~Bemfica, M.~M.~Disconzi and J.~Noronha,
Phys. Rev. D \textbf{98} (2018) no.10, 104064.

\bibitem{Bemfica:2019knx}
F.~S.~Bemfica, M.~M.~Disconzi and J.~Noronha,
Phys. Rev. D \textbf{100} (2019) no.10, 104020.

\bibitem{Kovtun:2019hdm}
P.~Kovtun,
JHEP \textbf{10} (2019), 034.

\bibitem{Hoult:2020eho}
R.~E.~Hoult and P.~Kovtun,
JHEP \textbf{06} (2020), 067.

\bibitem{Bemfica:2020zjp}
F.~S.~Bemfica, M.~M.~Disconzi and J.~Noronha,
[arXiv:2009.11388 [gr-qc]].

\bibitem{Hoult:2021gnb}
R.~E.~Hoult and P.~Kovtun,
[arXiv:2112.14042 [hep-th]].

\bibitem{Biswas:2022cla}
R.~Biswas, S.~Mitra and V.~Roy,
[arXiv:2202.08685 [nucl-th]].

\bibitem{Mitra:2021owk}
S.~Mitra,
Phys. Rev. C \textbf{105} (2022) no.1, 014902.

\bibitem{Dusling:2009df}
K.~Dusling, G.~D.~Moore and D.~Teaney,
Phys. Rev. C \textbf{81} (2010), 034907.

\bibitem{Teaney:2013gca} 
D.~Teaney and L.~Yan,
Phys. Rev. C \textbf{89} (2014) no.1, 014901.

\bibitem{Kurkela:2017xis}
A.~Kurkela and U.~A.~Wiedemann,
Eur. Phys. J. C \textbf{79} (2019) no.9, 776.

\bibitem{Mitra:2020gdk}
S.~Mitra,
Phys. Rev. C \textbf{103} (2021) no.1, 014905.

\bibitem{Rocha:2021zcw}
G.~S.~Rocha, G.~S.~Denicol and J.~Noronha,
Phys. Rev. Lett. \textbf{127} (2021) no.4, 042301.

\bibitem{Rocha:2021lze}
G.~S.~Rocha and G.~S.~Denicol,
Phys. Rev. D \textbf{104} (2021) no.9, 096016.

\bibitem{Romatschke:2009im}
P.~Romatschke,
Int. J. Mod. Phys. E \textbf{19} (2010), 1-53.

\bibitem{Degroot}
S.~R.~De Groot, W.~A.~Van Leeuwen and C.~G.~Van Weert,
{\it Relativistic Kinetic Theory, Principles And Applications}
(North-holland, Amsterdam, 1980).

\bibitem{Mitra:2015yaa}
S.~Mitra, U.~Gangopadhyaya and S.~Sarkar,
Phys. Rev. D \textbf{91} (2015) no.9, 094012.

\bibitem{Monnai:2012jc}
A.~Monnai,
Phys. Rev. C \textbf{86} (2012), 014908.



\end{thebibliography}
\end{document}